\documentclass[12pt,english]{article}
\pdfoutput=1
\usepackage[T1]{fontenc}
\usepackage{amssymb}
\usepackage{amsmath}
\usepackage{cite}
\usepackage{epsfig}
\usepackage[unicode=true,
 bookmarks=true,bookmarksnumbered=false,bookmarksopen=false,
 breaklinks=false,pdfborder={0 0 1},backref=false,colorlinks=true]
 {hyperref}
\usepackage{breakurl}
\usepackage[hang,flushmargin]{footmisc}
\usepackage{breakurl}
\usepackage{color}

\makeatletter
\def\simgt{\mathrel{\lower2.5pt\vbox{\lineskip=0pt\baselineskip=0pt
           \hbox{$>$}\hbox{$\sim$}}}}
\def\simlt{\mathrel{\lower2.5pt\vbox{\lineskip=0pt\baselineskip=0pt
           \hbox{$<$}\hbox{$\sim$}}}}
\makeatother

\newcommand{\be}{\begin{equation}}
\newcommand{\ee}{\end{equation}}
\newcommand{\bea}{\begin{eqnarray}}
\newcommand{\eea}{\end{eqnarray}}
\newcommand{\eq}[2]{\be\begin{aligned}#1 \label{#2}\end{aligned}\ee}

\newcommand{\Fig}[1]{Fig.~\ref{#1}}
\newcommand{\Eq}[1]{Eq.~\eqref{#1}}

\newcommand{\Sec}[1]{Sec.~\ref{#1}}

\hypersetup{citecolor=blue,linkcolor=blue,urlcolor=blue}

\begin{document}

\interfootnotelinepenalty=10000
\baselineskip=18pt

\hfill CALT-TH-2020-006

\vspace{2cm}
\thispagestyle{empty}
\begin{center}
{\LARGE \bf
Classical Gravitational Scattering  \\ \smallskip
at ${\cal O}(G^3)$  from Feynman Diagrams \\
}
\bigskip\vspace{1cm}{
{\large Clifford Cheung and Mikhail P. Solon}
} \\[7mm]
 {\it ${}^a$Walter Burke Institute for Theoretical Physics\\[-1mm]
    California Institute of Technology, Pasadena, CA 91125 } \let\thefootnote\relax\footnote{\noindent e-mail: \url{clifford.cheung@caltech.edu}, \url{mpsolon@gmail.com}} \\
 \end{center}
\bigskip
\centerline{\large\bf Abstract}
\begin{quote} \small
We perform a Feynman diagram calculation of the two-loop scattering amplitude for gravitationally interacting massive particles in the classical limit. Conveniently, we are able to sidestep the most taxing diagrams by exploiting the test-particle limit in which the system is fully characterized by a particle propagating in a Schwarzschild spacetime. We assume a general choice of graviton field basis and gauge fixing that contains as a subset the well-known deDonder gauge and its various cousins.  As a highly nontrivial consistency check, all gauge parameters evaporate from the final answer.  Moreover, our result exactly matches that of Bern {\it et al.}~\cite{3PM}, here verified up to sixth post-Newtonian order while also reproducing the same unique velocity resummation at third post-Minkowksian order.
\end{quote}

\setcounter{footnote}{0}

\newpage

\section{Introduction}
The breakthrough observation of gravitational waves at LIGO/Virgo~\cite{LIGO} has triggered immense interest in bridging developments from the modern scattering amplitudes program to the physics of gravitational waves.  Building on past work on the inspiral problem based on graviton effective field theory (EFT)~\cite{oldQFT} and matching to a classical potential~\cite{oldEFT,Neill:2013wsa,Vaidya:2014kza}, many developments have now emerged which exploit classic methods \cite{Damour:2017zjx,2PM,Matching,HBET} as well as recent amplitudes advances \cite{newAMP} to investigate systems with \cite{newSPIN} and without spin \cite{3PM, 3PMlong}.  

Already, these efforts have culminated in genuinely new results which have yet to be fully verified through the existing conventional methods, which include the effective one-body formalism~\cite{EOB}, numerical relativity~\cite{NR}, self-force formalism~\cite{self_force}, and analytic calculations in the post-Newtonian (PN)~\cite{PN}, post-Minkowskian (PM)~\cite{PM,Damour:2017zjx} and nonrelativistic general relativity~\cite{NRGR,EFT,Foffa:2019yfl} approaches.  In particular, the recent calculation of the conservative Hamiltonian for binary dynamics at 3PM order~\cite{3PM, 3PMlong} overlaps and agrees with existing results in the PN expansion but also encodes an infinitude of new higher order velocity corrections. This new 3PM calculation entails an intricate ``vertical pipeline'' of tools which span string theory, effective field theory, and orbital mechanics.

The 3PM calculation \cite{3PM, 3PMlong} centers on the scattering amplitude for two massive, spinless bodies interacting via Einstein gravity. The multiloop integrand for this process is built from tree amplitudes constructed via the double copy construction \cite{DoubleCopy} and then fused together via generalized unitarity \cite{GeneralizedUnitarity}.  The resulting object is then integrated through a battery of relativistic and nonrelativistic methods.  The latter approach hinges on a crucial split between potential and radiation graviton modes which was first systematized for the binary inspiral problem in a quantum field theoretic context in the pioneering work of~\cite{NRGR} (see~\cite{Foffa:2019yfl} for a full treatment of the conservative 4PN Lagrangian in this framework). Finally, by matching the resulting scattering amplitude to an EFT for the binary system, one extracts the conservative potential governing the inspiral~\cite{oldEFT,Neill:2013wsa,Vaidya:2014kza,2PM}. 

To date, there are now a number of works studying the implications of this 3PM result as well as its consistency.  These include the study of the effect of these new 3PM corrections on the binding energy of a binary inspiral in comparison with numerical relativity~\cite{Antonelli:2019ytb}.   Currently, the 5PN term of the 3PM result has been verified~\cite{Bini:2019nra}, while other methods for EFT matching have also been devised~\cite{Matching}.  The case of massless scattering has also received more recent attention with new computations in supergravity as well as in Einstein gravity~\cite{GravitonScattering,Abreu:2020lyk,Bern:2020gjj}. In particular, \cite{Bern:2020gjj} utilizes the first complete evaluation of the two-loop four-graviton scattering amplitude in~\cite{Abreu:2020lyk} and confirms from first principles the classic result for the massless deflection angle in pure Einstein gravity in~\cite{ACV}.  Notably, these explicit calculations are all inconsistent with the 3PM dynamics conjectured in~\cite{DamourRecent}. 

In this paper we perform an independent and comprehensive check of the 3PM results in \cite{3PM, 3PMlong} using age-old tools from the perturbative, quantum field theoretic description of gravitons coupled to massive scalars.  To begin, we compute the two-loop integrand associated with gravitational scattering using Feynman diagrams.  For maximal generality, we perform this calculation assuming an {\it arbitrary} choice of local graviton field basis and gauge fixing.  At two loops, the resulting Feynman diagrams individually depend on a total of ten gauge parameters, for which certain choices of values correspond to deDonder gauge, its nonlinear generalization to harmonic gauge, and a ``simplified'' gauge previously engineered to reduce the complexity of graviton perturbation theory~\cite{hidden}. Note that the latter formalism admits a version of Berends-Giele recursion relations \cite{BG} for gravity which was employed in the recent calculation of two-loop graviton scattering in Einstein gravity~\cite{Abreu:2020lyk}.

As a crucial simplification, we are able to completely sidestep a large class of Feynman diagrams which contribute only in the test-particle limit.  Instead, we fix these contributions from the known behavior of a test particle in a Schwarzschild background, as discussed in \cite{3PMlong}.

Afterwards, we integrate our two-loop integrands using the nonrelativistic method discussed in \cite{3PMlong} up to 6PN order.  As expected, all dependence on unphysical gauge parameters disappear entirely from our final result, which also exactly matches that of \cite{3PM}.
 As in~\cite{3PMlong}, the series of velocity corrections which appear are simple and take the same form as those collected in Appendix C of \cite{3PMlong}\footnote{Eq.~(C.4) of v2 of~\cite{3PMlong} has a typo. The right-hand side should be ${3E_1E_2 \over 2\boldsymbol{p}^2} \left[ 1 - {m_1^2 {\rm arcsinh} {|\boldsymbol{p}| \over m_1} + m_2^2 {\rm arcsinh} {|\boldsymbol{p}| \over m_2}  \over E |\boldsymbol{p}|} \right]$.}.  Resummation is then mechanical and reproduces the same 3PM scattering amplitude.  This agreement implies that the entire 3PM calculation---in particular the integration procedure---is gauge invariant, and furthermore that 
integrand construction via the double copy and generalized unitarity works as expected. Note that, along with the recent result for massless scattering at two loops~\cite{Bern:2020gjj}, this offers additional evidence via explicit computation that the claimed 3PM result in \cite{DamourRecent} is incorrect. Nonetheless, confirmation of the 6PN result from other methods, as was done at 5PN~\cite{Bini:2019nra}, would be valuable.\footnote{Note added: as this manuscript was in the final stages of preparation we became aware of~\cite{Blumlein:2020znm} which also confirms the 6PN result in~\cite{3PM} using the formalism of nonrelativistic general relativity.}

The paper is organized as follows. In \Sec{sec:action}, we discuss the action describing our setup for different choices of gauge fixing and field basis.  We then give a brief review of the classical limit of Feynman diagrams implemented at integrand level in \Sec{sec:classical}.  Afterwards, in \Sec{sec:testparticle} we discuss the subclass of Feynman diagrams that contribute only to the test-particle limit and show how to sidestep their direct calculation. In \Sec{sec:diagrams}, we list the final set of Feynman diagrams that we compute, and we discuss our results and outlook in \Sec{sec:results}.

\section{Setup}
The scattering amplitude for massive, gravitationally interacting particles is computed using the quantum field theory description of gravitons. For a review we refer the reader to~\cite{oldQFT}.  Here we specify various forms of the action that we use for our calculation.

\subsection{Action}\label{sec:action}

To begin, we consider Einstein gravity coupled to a pair of massive scalars.  The action is\footnote{We work in mostly plus metric signature throughout.}
\eq{
S &= S_{\rm graviton} + S_{\rm matter} + S_{\rm GF}
}{}
where the graviton and matter actions are
\eq{
S_{\rm graviton} &=  \frac{1}{16\pi G} \int d^Dx \sqrt{-g} \, R \\
S_{\rm matter} &=  \int d^Dx \sqrt{-g}  \sum_{i=1,2}  \left( -\frac{1}{2} \nabla_\mu \phi_i  \nabla^\mu \phi_i - \frac{1}{2} m_i^2\phi_i^2 \right)
}{eq:S}
and $S_{\rm GF}$ denotes the gauge fixing term.  Here all derivatives and metric contractions are covariant with respect to the full metric.

In order to define perturbation theory for the graviton, \Eq{eq:S} must be supplemented with an explicit definition of the graviton fluctuation about flat space.  As explored in \cite{twofold}, there is immense freedom in this choice of field basis and gauge fixing which will affect intermediate steps in any calculation but will evaporate from any physical quantity.  Let us describe the various choices of gauge fixing and field basis to be considered in this paper.

\medskip

\noindent {\bf deDonder Gauge.}  To begin, let us consider \Eq{eq:S} with a gauge fixing and field basis,
\eq{
S_{\rm GF} &= - \frac{1}{32\pi G} \int d^D x  \left(\partial^\nu h_{\mu\nu} -\frac{1}{2} \partial_\mu h\right)^2 \qquad {\rm and} \qquad g_{\mu\nu} = \eta_{\mu\nu} + h_{\mu\nu},
}{eq:LH}
where $h_{\mu\nu}$ is the graviton, $h$ is its trace, and all contractions are taken with the flat space metric $\eta_{\mu\nu}$.  Throughout, we work in a noncanonical normalization in which the graviton is dimensionless.  Since we will not be concerned with processes with external gravitons, this choice will not affect the overall normalization of the scattering amplitude.  That said, in order to go to canonical normalization, one simply rescales the graviton by a factor of $\sqrt{32\pi G}$. 

We emphasize that \Eq{eq:LH} is purposely expressed in terms of partial derivatives and does not include a factor of $\sqrt{-g}$.  This term is obviously not gauge invariant but this is expected since it is a gauge fixing term.  Since the gauge fixing term is purely quadratic in the graviton, it serves only to modify the graviton propagator of the theory.  Consequently, the Feynman rules for this formulation are obtained by inserting the definition of the graviton perturbation in \Eq{eq:LH} into the Einstein-Hilbert and matter coupling terms while using the well-known deDonder propagator for the graviton.

\medskip

\noindent {\bf Harmonic Gauge.} Alternatively, consider \Eq{eq:S} with the gauge fixing
\eq{
S_{\rm GF} &= - \frac{1}{32\pi G} \int d^D x \sqrt{-g}  \,  \Gamma^{\mu\nu}_{\;\;\;\;\nu}\Gamma_{\mu\rho}^{\;\;\;\;\rho}  \qquad {\rm and} \qquad g_{\mu\nu} = \eta_{\mu\nu} + h_{\mu\nu},
}{eq:NLH}
where the indices in the Christoffel symbols are contracted using the full metric.
Harmonic gauge fixing is a nonlinear generalization of the deDonder gauge in \Eq{eq:LH} since they coincide at quadratic order in the graviton but deviate at higher order.  In particular, graviton self-interactions arise from the gauge fixing term as well as the Einstein-Hilbert term.    However, harmonic gauge and deDonder gauge exactly coincide at quadratic order in the graviton, so the propagator here is still of deDonder form.

\medskip

\noindent {\bf Simplified Gauge.} In \cite{twofold,hidden}, the Einstein-Hilbert action was analyzed in an arbitrary field basis and gauge fixing.  While these choices have no effect on physical observables, they can elucidate various hidden structures in gravity and also simplify the Feynman diagram expansion.  For example, \cite{twofold} showed how the dual Lorentz invariance implied by the double copy construction \cite{DoubleCopy} can be made manifest at the level of the action.  

In \cite{hidden}, these freedoms were further exploited to build highly simplified Feynman rules and Berends-Giele recursion relations \cite{BG} for gravity derived from a perturbative version of the Palatini formulation.  In fact, this work was later utilized in the first calculation \cite{Abreu:2020lyk} of the two-loop scattering amplitude of massless gravitons.
In the present work, we also use this simplified action for graviton perturbations about flat space, $S = S_{\rm graviton} + S_{\rm matter}$, where
\eq{
S_{\rm grav} =  \frac{1}{16\pi G} & \int d^Dx   \left[ -\frac{1}{4} \partial_\mu h_{\nu\rho}\partial^\mu h^{\nu\rho} +\frac{1}{D-2} \partial_\mu h\partial^\mu h \right. \\
&  \left. + \frac{1}{4} h^{\mu\nu} \left( \partial_\mu h_{\rho \sigma} \partial_\nu h^{\rho\sigma}+
2\partial_{[\rho} h_{ \sigma]\nu} \partial^\sigma h_{\mu}^{\;\;\;\rho}
+\frac{1}{D-2} (2\partial_\rho h_{\mu\nu} \partial^\rho h-\partial_\mu h \partial_\nu h) \right) \right] +\cdots
}{}
where the ellipses denote terms which are quartic or higher in the graviton which are not needed for the present calculation.
As before, the graviton kinetic term again coincides with that of deDonder and harmonic gauge, so the graviton propagator is the same. 
Meanwhile, the graviton couplings to matter are slightly modified, and should be obtained by inserting the graviton perturbation defined by the relation $\sqrt{-g}\, g^{\mu\nu} = \eta^{\mu\nu} - h^{\mu\nu}$ into \Eq{eq:S}.

\medskip

\noindent {\bf Generalized Gauge.} Last but not least, we consider the Einstein-Hilbert action in a general field basis and gauge fixing, subject only to the assumption of locality.  In particular, we consider the most general local gauge fixing term that coincides with harmonic gauge at linear order while maintaining the deDonder form of the graviton propagator,
\eq{
S_{\rm GF} &= - \frac{1}{32\pi G}\int d^D x \,  F_{\mu}F^\mu 
}{}
where the gauge fixing function is
\eq{
F_\mu &= \partial^\nu h_{\mu\nu} -\frac{1}{2} \partial_\mu h +  \partial^\nu h^{\rho\sigma} ( \zeta_1 h_{\mu\nu}\eta_{\rho\sigma}+\zeta_2 h_{\mu\rho}\eta_{\nu\sigma}+ \zeta_3 h_{\rho\sigma}\eta_{\mu\nu}+\zeta_4 h_{\nu\sigma}\eta_{\mu\rho}+\zeta_5 \eta_{\mu\nu}\eta_{\rho\sigma}h +\zeta_6 \eta_{\mu\rho}\eta_{\nu\sigma}h  )  ,
}{eq:F_gen}
and all contractions are taken with the flat space metric. At linear order, this gauge fixing coincides with deDonder gauge.  
Furthermore, in the Einstein-Hilbert term we plug in the general graviton field basis,
\eq{
 g_{\mu\nu} &= \eta_{\mu\nu} + h_{\mu\nu}+ \xi_1 h_{\mu \rho} h_\nu^{\;\;\rho} + \xi_2  h_{\mu\nu}h  + \xi_3 \eta_{\mu\nu}  h_{\rho\sigma} h^{\rho\sigma}+ \xi_4 \eta_{\mu\nu}  h^2,
}{eq:basis_gen}
restricting to local nonlinear functions of the graviton.   In \Eq{eq:F_gen} and \Eq{eq:basis_gen} we have neglected to write down terms even higher order in the graviton since these only affect quartic or higher self-interaction vertices which we will not need.

Last of all, note that since the deDonder, harmonic, and simplified gauge discussed earlier are all local functions of the graviton field, they are all subsumed by various choices of the gauge parameters above.

\subsection{Classical Limit}\label{sec:classical}

As discussed at length in \cite{2PM,3PMlong}, the complexity of the scattering amplitude calculation is immensely reduced by the fact that we are interested only in the classical dynamics.  Indeed, the vasty majority of terms computed via Feynman diagrams actually contribute to the quantum dynamics.  By applying classical truncation as early as possible---in particular at the level of the integrand---we can substantially simplify our expressions.

Consider, for example, a scattering process with center of mass momentum $\vec p$ and momentum transfer $\vec q$.  Because the impact parameter scales as $b \sim 1/|\vec q|$, then the angular momentum of the process goes as $J \sim |\vec p| / |\vec q|$.    For a classical process, the angular momentum must be large in units of $\hbar$, so $J \gg 1$. Mechanically, we can enforce this hierarchy in Feynman diagrams by scaling relativistic momenta as follows
\eq{
p_{1} &\rightarrow  p_{1}  \\ 
p_{2} &\rightarrow  p_{2}  \\ 
q &\rightarrow \lambda q \\
 \ell_i &\rightarrow \lambda \ell_i,
}{eq:lambda_rescale}
where $p_{1}$ and $p_2$ are the incoming momenta of particles 1 and 2, respectively, $q=(0,\vec{q})$ is the momentum transfer in the center of mass frame, $\ell_i$ denotes the (loop) four-momenta of exchanged gravitons, and $\lambda$ is the classical power counting parameter. We expand in small $\lambda$ in the numerator. On the other hand, as discussed at length in \cite{3PMlong}, we do {\rm not} expand propagators in Feynman diagrams at this stage in order to keep the pole structure manifest for loop integration. For a detailed discussion of the nonrelativistic integration method, we refer the reader to \cite{3PMlong}.

With this power counting, the classical momentum-space scattering amplitude at $n$th order in the PM expansion scales as
\eq{
{\cal M}_{n} \rightarrow \lambda^{n-3} {\cal M}_n + \textrm{lower order in }\lambda.
}{eq:M_rescale}
Terms higher order than $\lambda^{n-3}$ are pure quantum and should be discarded.  Note, on the other hand, that there do in general exist ``super-classical'' terms which are lower order in $\lambda$, infrared divergent, and enter through iterations of lower orders in the PM potential.  While super-classical terms appear in the scattering amplitude, they evaporate from the conservative potential after matching, which is expected since the full theory the EFT share the same infrared structure.

As discussed in \cite{3PMlong}, the classical limit also permits us to drop large classes of Feynman diagrams. These diagrams can be dropped because their corresponding integrands have no poles in the integration region corresponding to potential graviton modes.  Consequently, they contribute only when the exchanged gravitons are high energy, thus generating pure quantum corrections.

\subsection{Test-Particle Limit}\label{sec:testparticle}

In the test-particle limit, $m_1 \gg m_2$, the scattering amplitude coincides with that of a point particle propagating in a Schwarzschild background. This well-known fact was  one of many consistency checks of the 3PM result~\cite{3PM,3PMlong}. In the present work we use this limit to sidestep the calculation of a complicated subset of Feynman diagrams. That is, we identify the class of Feynman diagrams that contribute only to the test-particle limit, and instead of computing them explicitly, fix them so that the final answer agrees with the test-particle limit.

\begin{figure}
\begin{center}
\includegraphics[scale=.48]{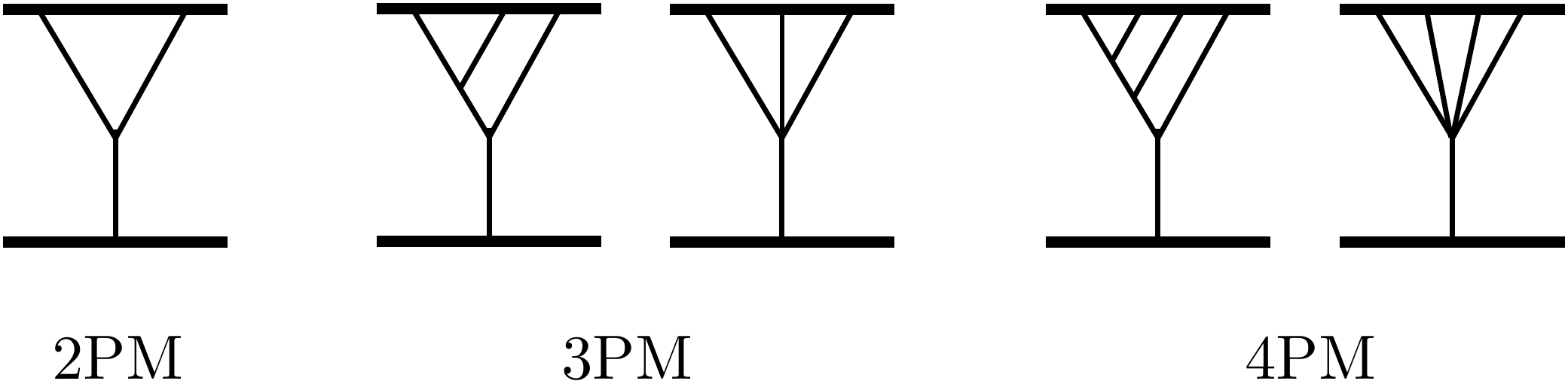}
\end{center}
\caption{Sample Feynman diagrams corresponding to the test-particle limit at 2PM, 3PM, and 4PM. Thick horizontal lines and thin lines respectively denote massive scalars and exchanged gravitons. Other variants include nonplanar topologies and those involving the seagull vertex.}
\label{fig:testparticle}
\end{figure}

Consider the ``triangular'' Feynman diagrams in \Fig{fig:testparticle}, which have the defining feature of containing the maximal allowed number of propagators for particle 1 but no propagators for particle 2. Crucially, this class of diagrams typically involves the highest order self-interactions of the graviton, and are very work-intensive to compute using Feynman diagrams.

First, let us apply to these diagrams the classical power counting outlined in \Eq{eq:lambda_rescale} and \Eq{eq:M_rescale}.  A diagram at $n$th PM order will involve an $n-1$ loop integral of the form
\eq{
{\cal I} \sim \int d^{4(n-1)}\ell \times \frac{1}{\ell^{n-1}} \times \frac{1}{\ell^{2n}} \times {\cal N},
}{eq:I_tri}
where ${\cal N}$ is the numerator of the Feynman diagram, and $\ell$ schematically denotes any loop momentum or momentum transfer, including $q$, which scales linearly with $\lambda$ in \Eq{eq:lambda_rescale}.  The $1/\ell^{n-1}$ and $1/\ell^{2n}$ come from the loop momentum dependence of the matter lines and the all-graviton sub-diagram, respectively.  Note that we have suppressed dependence on the scales in the problem, such as the masses.  Applying \Eq{eq:lambda_rescale} and comparing to  \Eq{eq:M_rescale}, it is then clear that we have to take ${\cal N}$ at zeroth order in the classical expansion parameter $\lambda$, thus setting everything in the numerator to zero except $p_1^\mu$, $p_2^\mu$, and the masses.

By this logic, the Feynman diagram numerator will carry $2(n+1)$ factors of  $p_1^\mu$ and two factors of $p_2^\mu$ after classical truncation, here rewriting all masses as $m_1^2 = p_1^2$ or $m_2^2 = p_2^2$ via the on-shell condition.   This implies that
\eq{
{\cal N} \sim G^n m_1^{2(n+1)} m_2^2 ( A + B \sigma^2),
}{}
where $\sigma = -{p_1 \cdot p_2  \over m_1m_2}$ in our mostly plus signature, and $A$ and $B$ are unknown dimensionless coefficients.  Higher powers of $\sigma$ cannot arise, simply because there are not enough factors of $p_2^\mu$ to produce them.  Furthermore, as described in \cite{3PMlong} the first step of integration is to localize all matter poles for particle 1, effectively introducing a total of $n-1$ factors of $1/m_1$.  Applying this to \Eq{eq:I_tri}, we obtain
\eq{
{\cal I} \sim   G^n m_1^{n+3} m_2^2 ( A + B \sigma^2)q^{n-3},
}{eq:gen_form}
where the power of $q=\sqrt{-t}$ is fixed by \Eq{eq:M_rescale} and is only schematic---that is, depending on the PM order, logarithms of $q$ may also appear.

Rather than compute the triangular Feynman diagrams explicitly, we simply add them to our calculation as an ansatz term in the general form of \Eq{eq:gen_form}.  Obviously, we can do the same for diagrams related to those in \Fig{fig:testparticle} by swapping particles 1 and 2, and they will have the same values for $A$ and $B$.  
In particular, for the 3PM answer we take
\eq{
{\cal M}_{3{\rm PM}} =  {\cal M}_{\rm ansatz} + {\cal M}_{\rm Feynman}
}{}
where the ansatz function is
\eq{
{\cal M}_{\rm ansatz} &= \pi G^3 m_1^2 m_2^2 (m_1^2+ m_2^2) (A+ B \sigma^2) \ln q,
}{eq:M_ansatz}
where the $q$ dependence is fixed by the $1/r^3$ structure of the 3PM potential.  The remaining term ${\cal M}_{\rm Feynman}$ is computed explicitly via Feynman diagrams.  We then take the full answer ${\cal M}_{3{\rm PM}}$ and demand that it is consistent with the test-particle limit amplitudes presented in~\cite{3PMlong}.  This constraint uniquely fixes $A$ and $B$, thus completing the calculation.

\subsection{Feynman Diagrams}\label{sec:diagrams}
\begin{figure}
\begin{center}
\includegraphics[scale=.48]{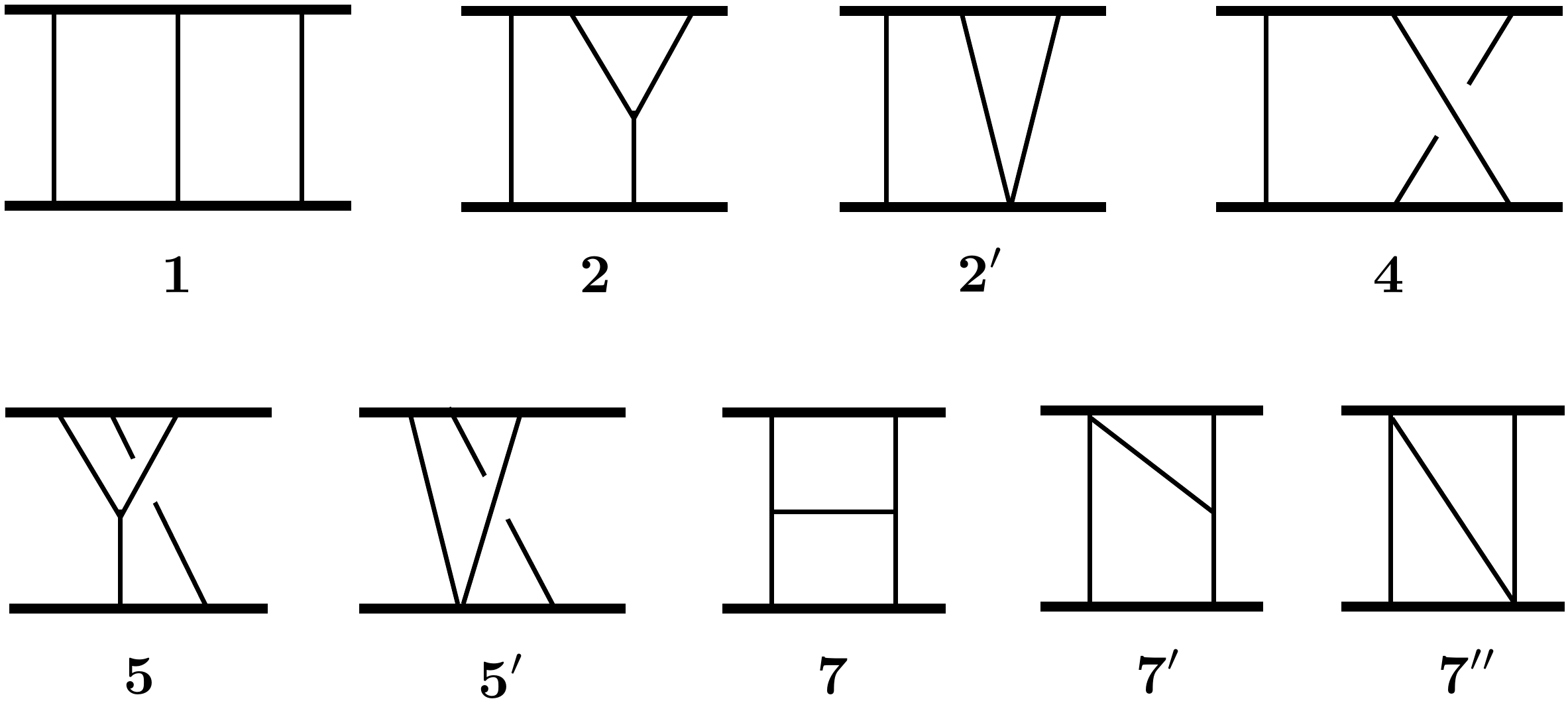}
\end{center}
\caption{Two-loop Feynman diagrams for classical scattering. Not shown here are diagrams such as those in \Fig{fig:testparticle}, which are trivially fixed by the test-particle limit, as well as ``twisted'' graphs obtained by swapping the incoming and outgoing legs for particle 1, or equivalently, for particle 2. The peculiar labeling is meant to align with the topologies defined in Fig.~14 of~\cite{3PM}, and the primed labels denote graphs in which an exchanged graviton has been pinched.}
\label{fig:diagrams}
\end{figure}

Upon dropping quantum and test-particle contributions, we need only compute the subset of Feynman diagrams shown in \Fig{fig:diagrams}.  Notably, these diagrams involve at most cubic self-interactions of the graviton but not higher, affording some degree of reduction in complexity.

For the interested reader, we include an attachment containing all two-loop integrands, classically truncated, for the deDonder, harmonic, and simplified gauges described above, as well as their integrated values. We include only the finite pieces, while the infrared divergent, super-classical terms that cancel in the matching with the EFT are as given in Eq.~(9.2) of~\cite{3PMlong}.

\section{Results and Outlook}\label{sec:results}

In summary, we have computed all of the Feynman diagrams in \Fig{fig:diagrams} working in the deDonder, harmonic, simplified, and general gauges discussed previously. Integration was performed using the nonrelativistic method discussed at length in~\cite{3PMlong}. 
Calculating up to 6PN order in the velocity expansion, we find perfect agreement among all four gauges and with the results of~\cite{3PM,3PMlong}.  For all cases, the test-particle limit fixes  $A=0$ and $B=-64$ in \Eq{eq:M_ansatz}. 
As discussed in~\cite{3PMlong}, due to the limited number of possible relativistic invariants, agreement at 6PN is sufficient to guarantee a simple and unique expression for the 3PM amplitude and potential.

The fact that these separate computations all yield the same answer is a highly nontrivial verification of our previous 3PM result. On the one hand, the present calculation is a test of the integrands previously computed via generalized unitarity.  Furthermore, this computation confirms that the nonrelativistic integration method devised in~\cite{2PM,3PMlong} is fully gauge invariant.   While the methods employed in~\cite{2PM,3PMlong} are well-established tools of effective field theory and scattering amplitudes,  it is nevertheless necessary to perform such checks in light of the doubts recently raised in \cite{DamourRecent}.

As we have seen, the 3PM calculation relevant to conservative binary dynamics is actually tractable via standard Feynman diagrammatic methods.  This simple fact strongly suggests that amplitudes methods---which are invariably more efficient than Feynman diagrams---will have mileage to even higher orders than expected.

 \begin{center} {\bf Acknowledgments}
\end{center}
\noindent 
We thank Zvi Bern, Aneesh Manohar, Ira Rothstein, and Nabha Shah for helpful discussions. C.C. and M.P.S. are supported by the DOE under grant no.~DE- SC0011632 and by the Walter Burke Institute for Theoretical Physics. The calculations here
used the computer algebra system \texttt{Mathematica}~\cite{Mathematica} in combination with \texttt{FeynCalc}~\cite{FeynCalc} and \texttt{xAct}~\cite{xAct}.


\end{document}